# Controlled plasmon-enhanced fluorescence by spherical microcavity

Jingyi Zhao(赵静怡),[1,‡] Weidong Zhang(张威东),[1,‡] Te Wen(温特),[1] Lulu Ye(叶璐璐),[1] Hai Lin(林海),[1] Jinglin Tang(唐靖霖),[1] Qihuang Gong(龚旗煌)[1,2,3] and Guowei Lu(吕国伟),[1, 2,3,†]

[1] *State Key Laboratory for Mesoscopic Physics, Frontiers Science Center for Nano-optoelectronics & Collaborative Innovation Center of Quantum Matter, School of Physics, Peking University, Beijing 100871, China*

[2] *Collaborative Innovation Center of Extreme Optics, Shanxi University, Taiyuan, Shanxi 030006, China*

[3] *Peking University Yangtze Delta Institute of Optoelectronics, Nantong 226010, Jiangsu, China*

A surrounding electromagnetic environment can engineer spontaneous emissions from quantum emitters through the Purcell effect. For instance, a plasmonic antenna can efficiently confine an electromagnetic field and enhance the fluorescent process. In this study, we demonstrate that a photonic microcavity can modulate plasmon-enhanced fluorescence by engineering the local electromagnetic environment. Consequently, we constructed a plasmon-enhanced emitter (PE-emitter), which comprised a nanorod and a nanodiamond, using the nanomanipulation technique. Furthermore, we controlled a polystyrene sphere approaching the PE-emitter and investigated *in situ* the associated fluorescent spectrum and lifetime. The emission of PE-emitter can be enhanced resonantly at the photonic modes as compared to that within the free spectral range. The spectral shape modulated by photonic modes is independent of the separation between Project supported by the National Key Research and Development Program of China (grant no. 2018YFB2200401), Guangdong Major Project of Basic and Applied Basic Research (No. 2020B0301030009), and the National Natural Science Foundation of China (grant nos. 91950111, 61521004, and 11527901).
‡ These authors contributed equally.
† Corresponding author. E-mail: guowei.lu@pku.edu.cn

the PS sphere and PE-emitter. The band integral of the fluorescence decay rate can be enhanced or suppressed after the PS sphere couples to the PE-emitters, depending on the coupling strength between the plasmonic antenna and the photonic cavity. These findings can be utilized in sensing and imaging applications.

**PACS:** 42.82.Fv (Hybrid systems) 73.20.Mf (Collective excitations (including excitons, polarons, plasmons and other charge-density excitations))

The approach of tailoring the spontaneous emission rate has made tremendous progress over the past few decades. [1, 2] Strong light-matter interactions facilitate efficient light guiding, [3-5] energy transfer, [6, 7] and control of emission properties at the scale of atoms, [8, 9] molecules, [10, 11] or quantum dots. [12, 13] A rich toolbox of photonic systems has been used to manipulate spontaneous emissions, such as those from microcavities, [14] photonic crystals, [15] and plasmonics. [16] According to Fermi's golden rule and the Purcell effect, [17] using optical resonators with high Q factors or small mode volumes (V) can engineer the electromagnetic environment and the photonic modes around the quantum emitters. Furthermore, it is well known that plasmonic materials and microcavities exhibit complementary optical properties. In particular, plasmonic devices offer strong optical confinement in the subwavelength regime and an ultrasmall mode volume (V$\sim \lambda^3/10^4$), but suffer from high Ohmic dissipation. In contrast, high-Q photonic cavities can sustain low-loss radiation; however, the field localization is inherently limited because of the relatively large mode volume.

In this case, the use of plasmonic materials together with microcavities seems to be more promising than solely utilizing a pure plasmonic system. Hybrid plasmonic-photonic resonators have been proposed for numerous applications, including biosensing, [18-22] light emission, [23-26] and nanoscale lasers. [27-29] For the hybrid plasmonic-photonic system, previous theoretical studies have suggested that hybrid modes offer Purcell factors that exceed individual constituents. Meanwhile, the presence of microcavities enhances the coherent radiation of the dipolar plasmonic mode, thus reducing incoherent Ohmic dissipation. [30-32] When interacting with a quantum emitter, the microcavity-engineered localized surface plasmon resonances (LSPRs) significantly enhance the quantum yield and the radiative power output as compared to those achieved in vacuum environments. When an antenna exhibits strong scattering coupling to a photonic cavity, emissions are suppressed since strong radiation damping reduces the polarizability of strong scatterers. [33] Recent experimental works have demonstrated the advantages of plasmonic-photonic hybrid cavity modes in spontaneous emission control. [34, 35] At room temperature, quantum emitters often present a broad spectrum, which contrasts to the linewidth of the photonic cavity mode. Understanding the fluorescence modulated by plasmonic-photonic hybrid cavity is helpful for surface-enhanced fluorescence-based applications. However, there is a lack of experimental investigations on the fluorescence modulations for such a broad spectral range.

In this work, we fabricated controllable hybrid resonators for modulating the emission spectrum of a single fluorescent nanodiamond (FND). We constructed a photonic-plasmonic hybrid structure that comprised a polystyrene (PS) sphere (~9.6 µm diameter) and a gold nanorod (GNR). The FND and GNR were assembled as a plasmon-enhanced emitter (PE-emitter). The distance between the PS sphere and PE-emitter could be controlled by using scanning probe manipulation. The fluorescent spectrum and lifetime of the FND coupled to the hybrid cavity mode can be measured *in situ*. The spectral shape reveals that the emission bands of the whispering gallery modes (WGMs) are enhanced as compared to those observed within the free spectral range of the PS sphere. The spectral shape modulated by photonic modes is independent of the distance between PS sphere and PE-emitter for the entire emission spectrum. Furthermore, the total fluorescent intensity in the broad emission band decreases for most cases after the PS sphere couples to the emitters. However, the emission can be enhanced resonantly at the WGMs with narrow bands. Therefore, we found that the spontaneous emission of the PE-emitter can be improved efficiently near the plasmonic resonant band by using WGM modes. The broadband integral of the fluorescence decay rate can be enhanced or suppressed after the PS sphere couples to the PE-emitters, depending on the coupling strength between plasmonic antenna and photonic cavity.

We constructed the hybrid system by using FND, GNR, and PS sphere in succession during the experiment. As shown in Figure 1(b), the atomic force microscope was first used for manipulating the GNR approaching the FND to form a PE-emitter. Owing to the coupling between the FND and the GNR, the PE-emitter retains its dipolar nature, thereby exhibiting a boosted decay rate. Then, a homemade fiber tip stuck to a tuning fork picks up the PS sphere from the substrate. We could easily manipulate the fiber tip to be above the PE-emitter by using XYZ-piezo. Furthermore, a plasmonic-WGM hybrid cavity could be created, as illustrated in Figure 1(c). An inverted microscope collects the fluorescent signal using a 60x/1.49 NA oil immersion objective lens. A continuous wavelength laser (at the wavelength of 532 nm, ~100 μW excitation power) and a picosecond laser (at the wavelength of 488 nm, ~50 μW excitation power) were used as the incident lasers for performing the fluorescent spectrum and lifetime measurements, respectively (Figure 1(d)). The fluorescent spectrum and lifetime before and after the coupling process can be measured *in situ* using a spectrometer and an avalanche photodiode through a TCSPC module.

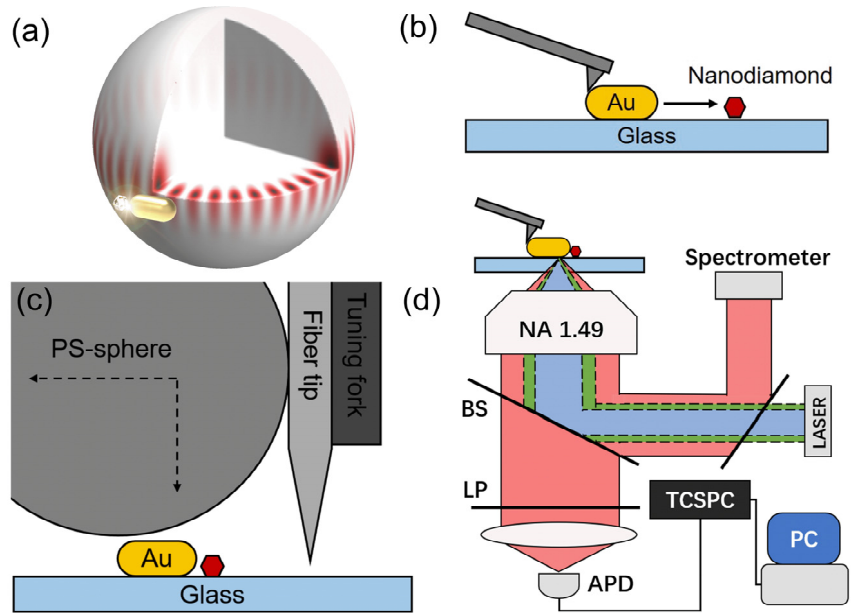

**Figure 1** (a) Hybrid system comprising a nanodiamond (FND) embedded in a photonic-plasmonic cavity formed by a PS sphere and a gold nanorod. (b) The AFM tip can precisely move the gold nanorod coupled with an FND under the semi-contact mode. (c) The homemade fiber tip sticks to the tuning fork, picks up the microsphere, and moves across the sample. The PS-sphere can be positioned above the PE-emitter using an *XYZ*-piezo. (d) Experimental setup: The sample is placed on an inverted microscope. The excitation laser (488 nm, 532 nm) focuses on the sample via the objective lens. A changeable mirror can switch the fluorescence to the spectrometer or the avalanche photodiode after being filtered by a long-pass filter (LP).

Before constructing the hybrid cavity, we first characterized the PE-emitter and the fluorescence of the FND modulated by the PS microcavity. As shown in Figure 2(a), the fluorescent peak of the bare FND is observed at approximately 670 nm. To achieve a higher enhancement factor, we used a GNR that exhibits resonance at the same emission wavelength of the FND. After coupling with the FND, the highly localized field near the GNR increases the absorption. Consequently, the fluorescence intensity can be enhanced considerably through the optical antenna effect. The PE-emitter emission spectra are modulated because the fluorescence signal from the FND scattered by the antenna is dominant. Moreover, the decay rate is measured simultaneously, as shown in Figure 2(b). The light emission lifetime of the GNR is too short to be resolved; therefore, it is treated as the instrument response function (IRF). To estimate the enhancement in fluorescence, we compared the lifetime of the PE-emitter to that of the bare FND. We can clearly observe the fluorescent lifetime of the PE-emitter becomes shorter from 20 ns to 10 ns.

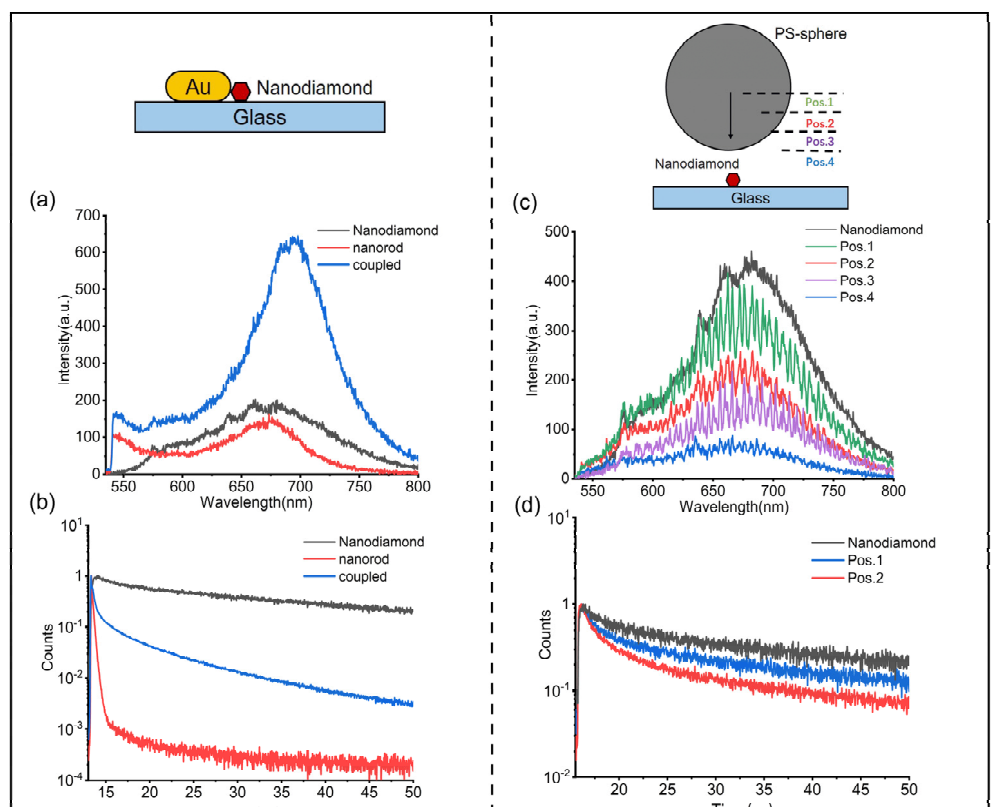

**Figure 2.** Nanodiamond (FND) coupled with a gold nanorod (left column) and a polystyrene sphere (right column). (a) Fluorescence spectra from the single nanorod (red), single FND (black), and coupled PE-emitter (blue). (b) Lifetime of the single FND (black), single gold nanorod (red), and the PE-emitter (blue). (c) Fluorescence spectra of a single FND (black), FND coupled with the microsphere at different positions (color). (d) The lifetime of the single FND (black), and that of the FND coupled with the microsphere at the far (blue) and near points (red).

Meanwhile, we used the fluorescence signal to characterize the WGM modes in the PS sphere. The PS sphere exhibits a diameter of 9.6 μm and is dispersed in aqueous solution. After being spin-coated onto the substrate, it can be picked up by a fiber tip. Then, the PS sphere is moved close to the FND on the glass substrate. The fluorescent spectrum shows many resonance peaks due to coupling with the PS microcavity (see Figure 2(c). We observed that the light emission spectral shape of the GNR can also be modulated using WGMs of the PS sphere. The Q factor (~ 3000) can be estimated through the linewidth of the sharp peaks associated with the WGM mode. Because the PS sphere is suspended in the air and we collected the fluorescent signal from the bottom, the far-field radiation is dissipated more in free spaces; therefore, the fluorescence intensity decreased after the coupling. Although the detected fluorescent intensity is not increased after coupling, we can characterize the total decay rate through the fluorescent lifetime. As shown in Figure 2(d), the fluorescent lifetime becomes shorter after coupling with the PS microsphere. The coupling between the FND and the microcavity WGM modes can accelerate the decay rate owing to the Purcell effect. We also find that the lifetime becomes shorter when the microsphere is closer to the FND.

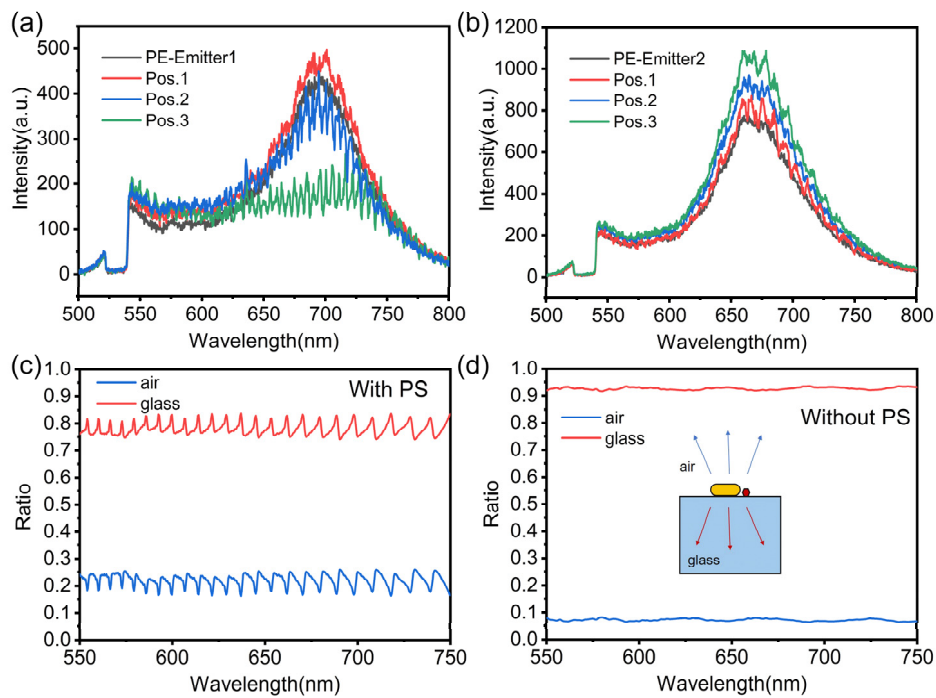

**Figure 3.** (a, b) Fluorescence spectra of the PE-emitter1&2 (black), coupled with the microsphere at different positions (color). (c, d) The ratio of the emission power toward air (glass) to the total emission power with and without PS.

We aimed to investigate how the PS sphere influences the fluorescence of the PE-emitter. As shown in Fig. 3(a) and (b), black lines represent the fluorescence spectra of the PE-emitter, which comprises an FND coupled with a GNR. When the microcavity approaches the PE-emitter with different positions, the detected fluorescence intensity of the PE-emitter either decreases or increases. We calculate the ratio of the emission power toward air or glass to the total emission power with and without the PS sphere, as shown in Fig. 3(c) and (d), respectively. After the PS sphere is coupled with the PE-emitter, the power ratio for the glass side decreased from 93% to 80%. The measured fluorescence intensity can be described as $I = \gamma_{exc} \cdot Q \cdot \varepsilon_{det}$, where $\gamma_{exc}$ is the excitation rate of the FND, Q is the fluorescent quantum efficiency, and $\varepsilon_{det}$ is the detection efficiency of the optical measurement equipment. Furthermore, the excitation rate and quantum efficiency of the PE-emitter can be modulated due to the engineered local electromagnetic environment provided by the PS microcavity. The large size of the microcavity may introduce more scattering of the light to the far-field, which implies that the detection efficiency could decrease after the PE-emitter couples to the PS sphere. Therefore, the fluorescent intensity is bound to decline in this situation. Overall, the competition among these factors determines the final fluorescent intensity.

An analogous situation to that mentioned above also occurs in the total decay rate of the hybrid system, as illustrated in Figure 4(c) and (e). Furthermore, it has been established that fluorescence emission can be enhanced or suppressed when a PE-emitter is placed near a microcavity, while an increase in the radiative decay rate and a reduction in the Ohmic dissipation contribute to enhancing the quantum yield of the PE-emitter. The coupling efficiency between the microcavity and the PE-emitter increases with a decrease in the distance between the microcavity and PE-emitter. Subsequently, the radiative decay rate is observed, which enhances the total decay rate of the hybrid structure (Figure 4(c)). When the coupling efficiency between the microcavity and PE-emitter is small, Ohmic dissipation dominates. As shown in Figure 4(e), when the PS sphere is close to the PE-emitter, the radiative decay rate is enhanced, and the Ohmic dissipation decay rate is reduced. In this stage, the total decay rate of the hybrid structure is reduced.

To realize the theoretical understanding of this phenomenon, we designed an analytical quantum model to describe the hybrid photonic-plasmonic mode; the Hamiltonian of the cavity mode c at $\omega_c$ is described as $H_c = \omega_c c^\dagger c$. Considering the FND to be a two-energy-level atom system, the Hamiltonian is written as $H_e = \omega_e \hat{\sigma}_z/2$, in which $\hat{\sigma}_z = |e><g|$ represents the transition operators. Furthermore, we consider that only dipolar LSPR couples to the microcavity; therefore, the plasmonic mode is written as $H_p = \omega_p \hat{a}_p^\dagger \hat{a}_p$. Accordingly, the free Hamiltonian of the whole system without any interaction is written as: $H_0 = H_c + H_p + H_e$; the interaction Hamiltonian is $H_I = G_{pc}(\hat{a}_p^\dagger \hat{c} + \hat{a}_p \hat{c}^\dagger) + G_{pe}(\hat{a}_p^\dagger \hat{\sigma}_- + \hat{a}_p \hat{\sigma}_+) + G_{ce}(\hat{c}^\dagger \hat{\sigma}_- + \hat{c} \hat{\sigma}_+)$. Thus, the quantum Langevin equations are given by

$$\frac{d\hat{c}}{dt} = -\left(i\omega_c + \frac{\kappa_c}{2}\right)\hat{c} - iG_{pc}\hat{a}_p - iG_{ce}\hat{\sigma}_-$$

$$\frac{d\hat{a}_p}{dt} = -\left(i\omega_p + \frac{\kappa_p}{2}\right)\hat{a}_p - iG_{pc}\hat{c} - iG_{pe}\hat{\sigma}_-$$

$$\frac{d\hat{\sigma}_-}{dt} = -\left(i\omega_e + \frac{\gamma_e}{2}\right)\hat{\sigma}_- - i\hat{\sigma}_z(G_{ce}\hat{c} + G_{pe}\hat{a}_p) - \sqrt{\gamma_{in}}\hat{\sigma}_{in,-}$$

where $\kappa_c$ and $\kappa_p$ represent the decay rates of the cavity mode and the dipolar plasmonic mode, respectively. $\gamma_e$ and $\omega_e$ denote the decay rate and the transition frequency of the emitter, respectively. $G_{pe}(G_{pc})$ indicates the coupling coefficient between the emitter (cavity) and the metal nanoparticle (MNP). $\hat{\sigma}_{in,-}$ represents the pump laser at the rate $\gamma_{in}$ and frequency $\omega_{pump}$. The interactions in the system are shown in Figure 4(a), in which $\kappa_p = \kappa_r + \kappa_0$, $\gamma_e = \gamma_s + \gamma_m$. $\kappa_r$ and $\kappa_0$ indicate the radiative rate and Ohmic loss rate of the MNP, respectively; $\gamma_s$ and $\gamma_m$ represent the radiation rate from the emitter to the environment and the dissipation rate to multipole plasmonic modes, respectively. We can calculate the radiation power $\Phi_r$ and the Ohmic loss power $\Phi_d$ by using the following equations:

$$\Phi_r = \langle(\sqrt{\kappa_r}\hat{a}_p^\dagger + \sqrt{\gamma_s}\hat{\sigma}_+)(\sqrt{\kappa_r}\hat{a}_p + \sqrt{\gamma_s}\hat{\sigma}_-)\rangle + \kappa_c\langle\hat{c}^\dagger\hat{c}\rangle$$

$$\Phi_d = \kappa_o\langle\hat{a}_p^\dagger\hat{a}_p\rangle + \gamma_m\langle\hat{\sigma}_+\hat{\sigma}_-\rangle$$

When the above-mentioned two terms are added, the total output power is obtained, as shown in Figure 4(d, f). Furthermore, when we assume that the emitter and the MNP exhibit resonance ($\omega_p = \omega_e = 1.8 eV$), the intrinsic quantum yields for the emitter and MNP are 75% and 1%, respectively. We established a series of cavity modes that functioned from 1.63 eV to 2.0 eV, with an interval of 0.03 eV; these modes were coupled with an emitter and a plasmonic mode (quality factor Q = 400). In our experiment, the microcavity is moved using the fiber tip; therefore, $G_{pc} = 0$ and $G_{pe}$ had a fixed value initially. When the cavity gets closer, $G_{pc}$ becomes larger; we plotted the total output power versus pump-emitter detuning $\Delta_{p,e}$ and selected two situations that correspond to the experimental results. Figure 4(d) shows that when $G_{pe}$= 52 meV and $G_{pc}$ = 60 meV, the integrated intensity is larger than $G_{pc}$=0 meV, which implies the total decay rate is enhanced after the cavity coupled with the PE-emitter corresponds to the situation shown in Figure 4(c). When the coupling between the cavity and the MNP is weak ($G_{pc}$=40meV), the integrated intensity is smaller than $G_{pc}$=0 meV, as shown in Figure 4(f); this indicates that the total decay rate decreased after the microcavity coupled (Figure 4(e)). We can also observe that when the single WGM mode is resonant with the emitter and the LSPR mode, the output power at the resonant band is always enhanced due to the microcavity and suppressed at the nonresonant band. Moreover, when the single WGM mode, emitter, and LSPR mode are not resonant, the enhancement becomes weak. When a series of WGM modes couple with the emitter and the LSPR mode, the output power at different bands can be enhanced or suppressed. In this experiment, the microcavity provides a series of different detuning modes. The modes associated with the PS microcavity influence the total output power, corresponding to the measured fluorescent intensity in the broad spectral range. Consequently, we established a series of modes to couple with the PE-emitter and calculated the integrated intensity with different $G_{pe}$ and $G_{pc}$ values (Figure 4(b)). The results imply that the total decay rate of the PE-emitter is dependent on its separation and coupling strength with the PS sphere.

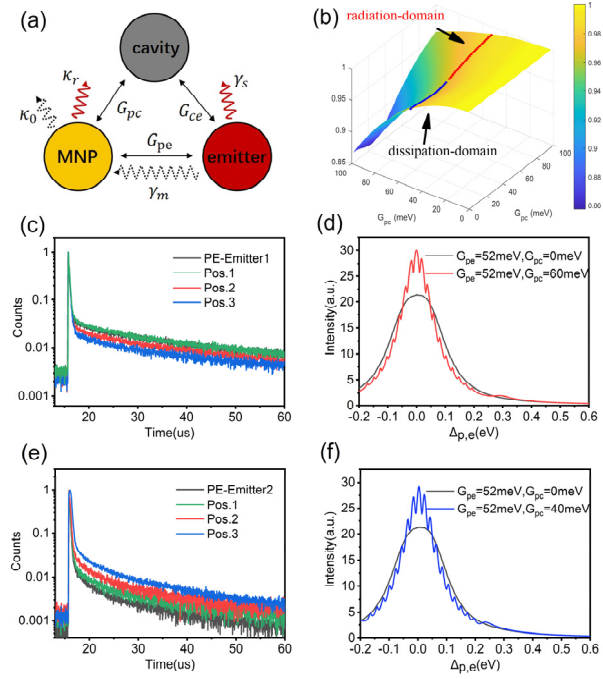

**Figure 4.** (a) Schematic of the interaction between the emitter, MNP, and microcavity. (b) Integrated intensity ($-0.15 < \Delta_{p,e} < 0.15$) versus $G_{pe}$ and $G_{pc}$. (c, e) Experimental result of different PE-emitters coupled with the PS sphere. (d, f) Theoretical calculation corresponding to the two different situations.

In summary, we have realized a controllable photonic-plasmonic hybrid system and observed the enhancement and suppression of the fluorescence emissions when an FND is placed near the hybrid structure. Furthermore, we observed that the PS photonic microcavity could control the plasmon-enhanced fluorescence. We were able to control a PS sphere approaching the PE-emitter and *in situ* performed fluorescent spectrum and lifetime measurements. The PE-emitter emission can be enhanced resonantly at the WGMs with a narrow band compared to that observed within the free spectral range of the PS sphere. The broadband integral of the fluorescence decay rate can be enhanced or suppressed after the PS sphere couples to the PE-emitters, depending on the coupling strength between the plasmonic antenna and the photonic cavity. Theoretical calculations imply that the microcavity cooperating with the plasmonic nanostructure modifies the density of optical states, leading to an enhancement of the radiative decay rate or suppression of the non-radiative rate. Our work also reveals the distance-dependent modulating behavior between the photonic microcavity and the plasmonic antenna. The hybrid system provides a novel approach to engineering the surrounding electromagnetic environment for controlling spontaneous emissions beyond the plasmonic nanostructure. For applications based on the hybrid system, strong interaction between the plasmonic antenna and the photonic cavity is essential to enhance the fluorescent intensity. More precise nanomanipulation will help to reveal the distance-dependent features of the hybrid system further.